\documentclass[12pt]{iopart}
\usepackage{graphicx}
\usepackage{caption}
\usepackage{booktabs}
\usepackage{bm}
\usepackage{color}
\usepackage{amssymb}%,,amsfonts
\begin{document}

\title[Identifying the structure patterns to govern the performance of localization]{Identifying the structure patterns to govern the performance of localization in regulating innovation diffusion}

\author{Leyang Xue$^{1}$, Peng-Bi Cui$^{1,*}$, Zengru Di$^{1}$}
\address{$^{1}$International Academic Center of Complex Systems, Beijing Normal University, Zhuhai, 519087, China}

\ead{cuisir610@gmail.com}

\vspace{10pt}
%\begin{indented}
%\item[]August 2017
%\end{indented}

\begin{abstract}
The macro social influence is recognized as a non-negligible ingredient in innovation propagation: more adopters in the network lead to a higher adoption tendency for the rest individuals. 
A recent study to incorporate such a crucial mechanism shows that sufficient intensity of macro-level social influence can cause a change from a continuous to discontinuous transition, further indicating the existence of a tricritical point.
Although network localization strength determines the tricritical point, it remains unclear what network quantities govern the performance of localization in regulating innovation diffusion. 
To address this issue, we herein consider the model incorporating both the micro- and macro-levels social influence. 
We present a dynamic message-passing method to analytically treat both the outbreak threshold and recovered population, and validate the predictions through agent-based simulations. 
Extensive analysis on the classical synthetic networks shows that sparsely available connections, and relatively heterogeneous degree distribution, either assortative or extremely disassortative configurations are favorable for continuous transition. 
In such cases, the employed network can yield a strong localization effect so that the innovation is trapped in the configurations composed of the hubs with high non-backtracking centrality. 
We further explore the dependence of both tricritical point and localization strength on three structural quantities: network density, heterogeneity, and assortativity, which gives a clear physical picture of the joint effects of the three structure quantities on the localization strength. 
Finally, we conclude that the core-periphery structure, being sensitive to the change of the three structure quantities, essentially determines localization strength, and further regulates the phase transition.
\end{abstract}

%
% Uncomment for keywords
%\vspace{2pc}
\noindent{\it Keywords}: complex network, contagion dynamics, macro-level social influence, phase transition, localization
%
% Uncomment for Submitted to journal title message
%\submitto{\JPA}
%
% Uncomment if a separate title page is required
%\maketitle
% 
% For two-column output uncomment the next line and choose [10pt] rather than [12pt] in the \documentclass declaration
%\ioptwocol
%

\section{Introduction}
\label{sec:introduction}
%describe the background of the innovation spreading
The propagation of innovation such as new ideas, technology, and products, is regarded as a fundamental process to drive the development of socioeconomic context~\cite{Dopfer2001}. However, it is quite difficult to exclude the effects of confounding factors in empirical studies. Therefore, how to model innovation diffusion becomes an increasing concern in understanding the underlying mechanisms of adoption decisions and guiding marketing strategies.
Extensive models have been proposed and studied from empirical and theoretical aspects.
Some studies have suggested that the payoff gain from innovation entities largely determines the final market penetration through shaping game profiles among individuals during the diffusion process~\cite{Young2009, Andrea2010, Gabriel2014}.
By means of a reaction-diffusion framework, others have highlighted the crucial role of local connectivity patterns of the target population in regulating the speed and extent with which innovation prevails~\cite{Castellano2009, Johan2012, Pastor2015,assenova2018}. For example, the hubs of the networks can largely facilitate the product~\cite{Pastor2015}. General mechanisms of social contagion have been additionally incorporated in the models belonging to this framework to enable a discontinuous transition, such as reinforcement~\cite{Centola2010,Guilbeault2018} and mutual cooperation between multiple diffusion process~\cite{Wang2019}. However, these research efforts mainly treat innovation diffusion as the context of social contagion~\cite{Castellano2009,Jusup2022}, calling for a fundamental understanding of network properties that regulate the innovation diffusion process. 

Recently, Ref.~\cite{Xue2022} has attempted to close this gap by proposing a model of innovation diffusion based on the susceptible-adopted-recovered (SAR) model, which incorporates social influence at micro and macro levels. The micro-level influence is conveyed through pairwise transmission between neighbors~\cite{Hohnisch2008}, representing the well-documented word-of-mouth effect~\cite{Buttle1998}.
While the macro-level influence promotes individuals' preference for adopting the new product as market penetration increases.
Such influence has at least three roots which innovation propagation is deeply involved: (i) People become more willing to follow the social norm with more adopters in the population, as a result of psychological effect such as group conformity~\cite{Efferson2008,Kandler2009,Young2009,Peres2010,Bale2013,Waal2013,Mccullen2013,Mccoy2014,Chen2016,Guilbeault2018}; (ii) The utility of the product increase due to network externalities (e.g. telecommunication products like fax, email, SNS and smartphone, new energy vehicles and artificial intelligence technology)~\cite{Katz1992,Schoder2000,Dopfer2001,Henrich2001,Rohlfs2003,Whiten2005,Young2015};
(iii) The product price might decrease with the marginal cost decline induced by economies of scale~\cite{Kiesling2012,Scherer1990,Browning2020}.
Both types of social influences together are assumed to contribute to the transmission, such that the transmissibility is a monotonic and increasing function of the recovered population. 
In Ref.~\cite{Xue2022}, the authors study the model on a collection of real-world networks and observe that the macro-level influence, depending on the proportion of the recovered population, can cause a change of the transition from continuous to discontinuous. 
There exist a tricritical point that separates the continuous and discontinuous phase transitions. 
They further focused on how the type of transition depends on the contact network topology by identifying the dependence of the tricritical points on the heterogeneity of non-backtracking centrality distribution. 
More in detail they argue that a discontinuous transition is more likely to occur if the network is dense and the high-centrality nodes are well separated, i.e., the localization effect is weak.
From this, they developed a novel metric to measure localization strength. 
Using the metric, one can accurately predict the position of a tricritical point on a large collection of empirical networks. 
Moreover, the metric can deeply reveal the relation between the critical and tricritical points, which further indicates a trade-off: networks that allow less attractive products to prevail instead tend to yield slower diffusion and lower market penetration, and verse versa.

An open question left by the study in Ref.~\cite{Xue2022} has to do with what structure patterns or quantities determine the performance of localization in regulating innovation diffusion with macro-level social influence. Simulations performed on a large collection of empirical networks seem to indicate that the tricritical point positively depends on the heterogeneity of non-backtracking centrality distribution. However, it is not clear what structure patterns affect the distribution and what happens for innovation diffusion with macro-level social influence if we tune these structure patterns. A clear physical picture of why the localization strength of the network can largely determine the magnitude of the tricritical point is still not integrated.

%In this work, describe what we have done.
In this paper, to address these questions, we systematically explore how classical synthetic networks govern the dynamics of innovation diffusion ruled by the model of Ref~\cite{Xue2022}. 
We herein use the dynamic message-passing (DMP) method to analytically track both the outbreak threshold and the final recovered population of the system. 
Numerical simulations validate the analytical predictions.
We perform the study on the following classical synthetic networks: two-dimensional~(2D), Erd\"{o}s-R\'{e}nyi~(ER), Watts–Strogatz~(WS), Barabasi-Albert~(BA), and scale-free~(SF) networks generated according to the uncorrelated configuration model~\cite{Catanzaro2005}.
As expected, with a given intensity of macro social influence, the innovation diffusion on the first four networks gives rise to discontinuous transitions. 
Moreover, the discontinuous ones become more apparent with denser connections. 
Weird contrary phenomena to the above cases are obtained for SF networks, changing degree exponent leads to a highly nonlinear change of phase transition. 
In more detail, continuous phase transitions can still take place when the network is sparse, assortative, or has a
relative heterogeneous degree distribution, reflecting the crucial roles of the spatial configurations of nodes, especially hubs. 
Furthermore, the effects of node configurations on the nature of phase transition are then specifically elucidated by varying the assortativity of SF networks. 
The likelihood of discontinuous transition decrease with network assortativity. Using the metric developed in Ref.~\cite{Xue2022}, we then show the dependence of both the tricritical point and localization strength on the three structure quantities. 
The dependent behaviors  surprisingly coincide with each other. 
Therefore, it is claimed that network density, heterogeneity, and assortativity are the structure quantities we are seeking for. 
Finally, we point out that core-periphery structure, being dependent on the three structure quantities, essentially determines localization strength, and further regulates the phase transition. 

%framework
The paper is organized as follows.
In Sec.~\ref{subsec:model}, we formally describe the innovation diffusion model incorporating macro-level social influence, followed by theoretical analyses in Sec.~\ref{subsec:analysis}.
In Sec.~\ref{subsec:structure}, we simulate our model on various networks and identify the key network structure characteristics that have a strong influence on the dynamical process.
Then a comprehensive analysis of the localization effect and its relation with the tricritical point is provided in Sec.~\ref{subsec:picture}.
We conclude with a summary of the results and an outlook for future studies in Sec.~\ref{sec:conclusion}.

\section{Method}
\label{sec:model_analysis}

\subsection{Model}
\label{subsec:model}

We consider an innovation diffusion model characterized by a social contagion process with a macro-level social influence.
Specifically, we use a SIR-like model to describe the contagion process, where an individual can be in one of three different states, i.e susceptible (denoted by $S$), adopted (denoted by $A$), or recovered (denoted by $R$).
In a discrete time step, the adopted individuals are enthusiastic about transmitting the contagion to their susceptible neighbors.
After interacting with an adopted neighbor, a susceptible individual may become an adopter as well with a probability $\beta$.
The adopted individuals may lose interest in advertising and become recovered (exhausted) with a probability $\mu$.
Moreover, the contagion process takes place on a network with $N$ nodes and $L$ edges.

To capture the individual's tendency from social influence, the transmissibility $\beta$ has two components, a constant part, and a time-dependent part, which is defined as
\begin{eqnarray}
	\label{eq:eqs3}
	\beta(t) =  \min(1,\beta_0 + \alpha \frac{R(t-1)}{N}), \alpha \geq 0, \beta_0 \in \left[ 0,1\right],  
\end{eqnarray}
where $\beta_0$ model the influence of direct and indirect biased transmission, which is the basic diffusion force from intrinsic attractiveness of the new entity, influenced by relative advantage, economic advantage, perceived compatibility, complexity, observability and so on~\cite{Rogers2003}. 
The second term $\alpha\frac{R(t-1)}{N}$ represents the contribution from the macro-level social influence, regarded as positive feedback from the macro-level back to the micro-level.
The parameter $\alpha$ measures the strength of influence, and $R(t-1)$ is the number of recovered individuals at time $t-1$. 
By varying the value of $\alpha$, we can control the intensity of macro-level social influence.

Eq.~\ref{eq:eqs3} indicates our assumption that the size of recovered population are transparent to the others. It is reasonable.
Since we focus on innovation diffusion, the market penetration of a product or a new technology is usually widely reported in the mass media and on the Internet. Our assumption is also supported by a wide array of existing literature on conformity effect~\cite{Mccoy2014,Young2009,Chen2016,Guilbeault2018}, network externalities~\cite{Katz1992,Schoder2000,Rohlfs2003}, and decline of marginal cost due to economies of scale~\cite{Kiesling2012,Scherer1990,Browning2020}.

With the model formally defined, we can simulate it on different network structures.
In each simulation, a node is randomly assigned the $A$ state to initiate the diffusion, while the remaining nodes are in the $S$ state.
The contagion process proceeds following the rules described above until an absorbing state is reached where no changes can be observed anymore.
Throughout this paper, the recovery probability $\mu$ is set to $1$.

\subsection{Theoretical Analysis}
\label{subsec:analysis}

In this section, we employ the DMP method to characterize the outbreak size and threshold analytically.
DMP prevents the contagion from backtracking to the node that it comes from, avoiding the mutual transmission effect~\cite{Karrer2010,lokhov2014,Lokhov2015,Koher2019,Li2021}.
Regardless of the initial conditions, DMP can accurately approximate the probability of each node being in a specific state at time $t$, especially for tree-like networks.

\subsubsection{Dynamic message-passing}
\label{subsubsec:dmp}
First, let us derive the exact equations of DMP.
The probabilities of node $i$ being in $S$, $A$, or $R$ states at time $t$ are represented by $P^i_{S}(t)$, $P^i_{A}(t)$ or, $P^i_{R}(t)$, respectively, with the following constraint:
\begin{equation}\label{eq:sum_to_one}
	P_A^i(t) + P_S^i(t) + P_R^i(t) = 1.
\end{equation}

The recovery process for an adopted node $i$ does not depend on its neighbors' state, so $P^i_{R}(t)$ can be represented as
\begin{equation}\label{eq:eqs4}
	P_R^i(t) = P_R^i(t-1) + \mu P_A^i(t-1).
\end{equation}
Subsequently, the global quantity $R(t)$ can be obtained using
\begin{equation}\label{eq:eqs5}
	R(t) = \sum_i^N{P_R^i(t)}.
\end{equation}
Substituting Eq.~(\ref{eq:eqs5}) into Eq.~(\ref{eq:eqs3}), we have
\begin{equation}\label{eq:eqs7}
	\beta(t) = min(1,\beta_0 + \alpha\frac{\sum^{N}_{i=1}P^{i}_{R}(t-1)}{N}),
\end{equation}
which increases with $R(t-1)$ until it reaches the upper bound $\beta(t) = 1$.

Next, we derive the expression for $P^i_{S}(t)$.
If we use $\Phi_i(t)$ to denote the probability that the contagion has not been successfully transmitted to node $i$ from its adopted neighbors until time $t$, $P^i_{S}(t)$ can be represented as
\begin{equation}\label{eq:eqs8}
	P_S^i(t) =  P_S^i(0) \Phi_i(t).
\end{equation}
DMP assumes that the underlying network is a tree and that the state evolution of node $i$'s neighbors is independent.
However, once node $i$ is successfully persuaded by its neighbor $z$ and switches to $A$ state, it would attempt to pass the contagion to another neighbor $z'$ ($z' \neq z$).
It gives rise to a dilemma because the state transitions of node $z'$ and $z$ are clearly correlated.
To alleviate this issue, we assume that the focal node $i$ is a cavity and define $\theta^{z\rightarrow i}(t)$ as the probability that node $i$ has not been successfully persuaded by node $z$ until time $t$.
$\Phi_i(t)$ can then be factorized as $\prod_{z\in \partial i} \theta^{z\rightarrow i}(t)$ where $\partial i$ is the neighbor set of node $i$.
Substituting it into Eq.~(\ref{eq:eqs8}), we have
\begin{equation}\label{eq:eqs9}
	P_S^i(t) = P_S^i(0) \prod_{z\in \partial i} \theta^{z\rightarrow i}(t). 
\end{equation}

Message-passing is directional, meaning that $\theta^{i\rightarrow z}(t) \neq \theta^{z\rightarrow i}(t)$ for undirected networks.
In our analysis, we treat each undirected edge as two directed edges pointing to opposite directions.
Initially, we have $\theta^{z \rightarrow i}(0) = 1$ for all edges in the network.
Later, $\theta^{z \rightarrow i}(t-1)$ decreases as the contagion transmits from node $z$ to $i$, which occurs with the probability $\beta(t)\phi^{z\rightarrow i}(t-1)$ where $\phi^{z\rightarrow i}(t-1)$ is the probability that the adjacent adopted node $z$ has not passed the contagion to node $i$ until time $t-1$.
Hence, $\theta^{z \rightarrow i}(t)$ follows the updating rule:
\begin{equation}\label{eq:eqs10}
	\theta^{z\rightarrow i}(t) = \theta^{z\rightarrow i}(t-1) - \beta(t) \phi^{z\rightarrow i}(t-1).
\end{equation}

Now we need to find the expression for $\phi^{z \rightarrow i}(t)$.
On the one hand, $\phi^{z \rightarrow i}(t)$ decreases when node $z$ in $A$ state recovers, or when the contagion is successfully transmitted from $z$ to $i$, or when the two processes occur simultaneously.
The corresponding probabilities for these events are $\mu$, $\beta(t)$, and $\mu \beta(t)$, respectively.
On the other hand, $\phi^{z \rightarrow i}(t)$ increases when node $z$ in $S$ state becomes $A$.
The changing rate $\Delta P^{z\rightarrow i}_S(t)$ can be calculated as $P_S^{z\rightarrow i}(t-1) - P_S^{z\rightarrow i}(t)$ where $P_S^{z\rightarrow i}(t)$ represents the probability of node $z$ staying in $S$ state after interacting with the cavity node $i$.
By combining all terms we have
\begin{eqnarray}\label{eq:eqs11}
    \phi^{z\rightarrow i}(t) & =  \phi^{z\rightarrow i}(t-1) - \beta(t)\phi^{z\rightarrow i}(t-1) -\mu\phi^{z\rightarrow i}(t-1) \\ \nonumber
      & + \mu\beta(t)\phi^{z\rightarrow i}(t-1) + P_S^{z\rightarrow i}(t-1) - P_S^{z\rightarrow i}(t)	\\ 
     & =  (1-\beta(t))(1-\mu)\phi^{z\rightarrow i}(t-1) + P_S^{z\rightarrow i}(t-1) - P_S^{z\rightarrow i}(t). \nonumber
\end{eqnarray}

The next step is to express $P_S^{z \rightarrow i}(t)$ explicitly.
Since node $i$ is a cavity, node $z$ will stay susceptible as long as it does not get the contagion from another neighbor $j$.
Using Eq.~(\ref{eq:eqs9}), we can obtain the probability that $z$ remains susceptible when its neighbor $i$ is a cavity: 
\begin{equation}\label{eq:eqs12}
	P_S^{z \rightarrow i}(t) = P_S^z(0) \prod_{j\in\partial z\backslash i} \theta^{j\rightarrow z}(t),
\end{equation}
where $\partial z\backslash i$ represents the neighbors of $z$ except for $i$.
At $t=0$, we have $P_S^{z\rightarrow i}(0)=0$ if $z$ is the seed that initiates the contagion process and $P_S^{z\rightarrow i}(0)=1$ otherwise.
This can be represented as $P_S^{z\rightarrow i}(0) = 1-\delta_{q_z(0),A}$ where $\delta_{q_z(0),A}$ is the Kronecker function and $q_z(0)$ represents the state of $z$ at time $t=0$.

Finally, we can employ Eqs.~(\ref{eq:eqs7}), (\ref{eq:eqs10})-(\ref{eq:eqs12}) to track the exact trajectory of $\theta^{z \rightarrow i}(t)$, $\phi^{z \rightarrow i}(t)$, and $P_S^{z \rightarrow i}(t)$ with the following initial conditions:
\begin{equation}\label{eq:eqs13}
    \theta^{z\rightarrow i}(0)  =   1, 
\end{equation}
\begin{equation}
\phi^{z\rightarrow i}(0) = P^z_A(0) = \delta_{q_z(0),A},
\end{equation}
\begin{equation}
   P^{z\rightarrow i}_S(0)  =  P^z_S(0) =  1-\delta_{q_z(0),A}.  
\end{equation}
By further combining $P_R^i(0)=0$ and $\beta(0)=\beta_0$ with Eqs.~(\ref{eq:sum_to_one}), (\ref{eq:eqs4}), (\ref{eq:eqs7}), and (\ref{eq:eqs9})-(\ref{eq:eqs12}) we can calculate the trajectory of $P^i_{S}(t)$, $P^i_{A}(t)$, $P^i_{R}(t)$, and obtain the order parameter $R(\infty)$.
The computation complexity of DMP is $O(L)$.

\subsubsection{Outbreak Threshold}
\label{subsubsec:threshold}
To obtain the outbreak threshold, we exert a small perturbation on the adopter-free state of the system and analyze its linear stability~\cite{Koher2019,Li2021}.
If the transmissibility reaches the threshold, the perturbation will lead to a global outbreak.
At the initial stage, we introduce two perturbation quantities $\epsilon_i \ll 1$ and $\delta^{z\rightarrow i}(t) \ll 1$ to Eqs.~(\ref{eq:eqs9}) and (\ref{eq:eqs10}) and express it as the following forms:
\begin{equation} \label{eq:eqs141}
    P^i_S(0)   =   1-\epsilon_i,
\end{equation}
\begin{equation}\label{eq:eqs142}
    \theta^{z\rightarrow i}(t) =  1 - \delta^{z \rightarrow i}(t). 
\end{equation}
Equation~(\ref{eq:eqs142}) and Eq.~(\ref{eq:eqs10}) together imply that 
\begin{equation}\label{eq:delta_updating_rule}
	\delta^{z \rightarrow i}(t) = \delta^{z \rightarrow i}(t-1) + \beta(t)\phi^{z\rightarrow i}(t-1).
\end{equation}

By substituting Eqs.~(\ref{eq:eqs141}) and (\ref{eq:eqs142}) into Eq.~(\ref{eq:eqs12}) we get
\begin{equation}
	P^{z\rightarrow i}_S(t) = (1-\epsilon_z)\prod_{j\in\partial z\backslash i} (1-\delta^{j\rightarrow z}(t)),
\end{equation}
and apply the approximation:
\begin{equation}\label{eq:p_approx}
	P^{z\rightarrow i}_S(t) \approx 1-\epsilon_z-\sum_{j\in\partial z\backslash i} \delta^{j\rightarrow z}(t).
\end{equation}
By substituting Eq.~(\ref{eq:delta_updating_rule}) into Eq.~(\ref{eq:p_approx}) we get
\begin{eqnarray}\label{eq:eqs16}
    P^{z\rightarrow i}_S(t)  & =  1-\epsilon_z-\sum_{j\in\partial z\backslash i} \delta^{j\rightarrow z}(t-1)  -\beta(t)\sum_{j\in\partial z\backslash i}{\phi^{j\rightarrow z}(t-1)}, \\ 
    & =  P_S^{z\rightarrow i}(t-1)- \beta(t)\sum_{j\in\partial z\backslash i}{\phi^{j\rightarrow z}(t-1)}, \nonumber 
\end{eqnarray}
where we apply the approximation in Eq.~(\ref{eq:p_approx}) again for $P_S^{z\rightarrow i}(t-1)$.
We further incorporate Eqs.~(\ref{eq:eqs16}) back into Eq.~(\ref{eq:eqs11}) and obtain the approximation of $\phi^{z\rightarrow i}(t)$:
\begin{equation}\label{eq:eqs17}
	\phi^{z\rightarrow i}(t) \approx (1-\beta(t))(1-\mu)\phi^{z\rightarrow i}(t-1) 
	+ \beta(t)\sum_{j\in \partial z\backslash i} {\phi^{j\rightarrow z}(t-1)}, 
\end{equation}
which is a self-consistent equation.
%$\beta(t)$ can now be expressed as
%\begin{eqnarray}\label{eq:eqs18}
%\beta(t) & = & \beta_0 + \frac{\alpha}{N}\sum_{i=1}^{N}(1-\mu)P_R^i(t-1)+\mu(1-P_S^i(t-1)),\notag \\
%& = & \beta_0 + \frac{\alpha}{N}\sum_{i=1}^{N} \sum_{n=0}^{t-1}{\mu(1-\mu)^n[1-P^i_S(t-1-n)]}.
%\end{eqnarray}

To track the dynamical trajectory of the whole system, we express Eq.~(\ref{eq:eqs17}) in matrix format as
\begin{equation}\label{eq:eqs20}
	\bm{\Phi}(t) = [(1-\mu)(1-\beta(t))\bm{I} +\beta(t)\bm{B}]\bm{\Phi}(t-1),
\end{equation}%e_{zi}=1, $\{\phi^{z\rightarrow i}(t):z, i \in \{1, 2, ..., N\}\}$
where $\bm{\Phi}(t)$ represents a vector with $\phi^{z\rightarrow i}(t)$ as the elements, $\bm{I}$ is a $2L\times 2L$ identity matrix, and $\bm{B}$ is a $2L\times 2L$ non-backtracking matrix with elements
\begin{equation}\label{eq:eqs19}
  	B_{z \rightarrow i,j\rightarrow z'} = \left \{
	\begin{array}{cc}
	1   \qquad   &  if \quad z' = z, j \neq i ,\\
	0   \qquad   &  otherwise. \\
	\end{array}
	\right.  
\end{equation}
For simplicity, we denote 
\begin{equation}
	\bm{\Phi}(t) = \bm{C}(t)\bm{\Phi}(t-1),
\end{equation}
and use a time-dependent propagator $\bm{P}$ to represent the accumulated effects of $\bm{C}(t)$:
\begin{equation}\label{eq:eqs21}
	\bm{P}(T,\beta_0,\mu,\alpha)=\prod_{t=1}^{T}\bm{C}(t),
\end{equation}
and further approximate Eq.~(\ref{eq:eqs20}) as 
\begin{equation}\label{eq:eqs211}
	\bm{\Phi}(T) = \bm{P}\bm{\Phi}(0).
\end{equation}
However, it is infeasible to qualitatively evaluate the asymptotic behavior of the dynamics through the spectral radius of $\bm{P}$ since it depends on time $t$.

To circumvent this obstacle, we define $\beta_c$ as the critical threshold.
If $\beta_0 < \beta_c$, the contagion remains confined to a small set of nodes and we have $\frac{R(t)}{N} \approx 0$ in the thermodynamic limit.
As a result, we have
\begin{equation}\label{eq:eqs212}
	{\lim_{N \to +\infty}} \beta(t \rightarrow \infty) = \beta_0 + \frac{\alpha R(t \rightarrow \infty)}{N} = \beta_0, 
\end{equation}
meaning that macro-level social influence is absent in this case. 
Now consider the scenario of  $\beta_0 > \beta_c$ where nodes in $R$ state will accumulate in the system and further amplify the positive feedback.
It suggests macro-level social influence only plays a critical role above $\beta_c$.
The position of phase transition does not change but is independent of $\alpha$, and can thus replace $\beta(t)$ with $\beta_0$ in Eq.~(\ref{eq:eqs20}). As a result, we have $\bm{C}=[(1-\mu)(1-\beta_0)\bm{I}+\beta_0\bm{B}]$ and $\bm{P} = \prod_{t=1}^T\bm{C}$ and it is possible to evaluate its asymptotic behavior through the spectral radius of $\bm{C}$, i.e., $\rho(\bm{C})$.
When the phase transition occurs, we have  
\begin{equation}\label{eq:eqs22}
	\rho([(1-\mu)(1-\beta_0)\bm{I}+\beta_0 \bm{B}])=1.
\end{equation}
Given that $0 \leq \beta_0, \mu \leq 1$ and all nodes in the network belong to a strongly giant connected component, $\bm{C}$ is non-negative and irreducible.
According to the Perron–Frobenius theorem~\cite{horn2012,lemmens2012}, we have $\lambda_C^{max}=\rho(\bm{C})$, $\lambda_C^{max}>0$, and there exists an positive eigenvector \bm{$v$} corresponding to $\lambda_C^{max}$.
With these relations, we can further simplify Eq.~(\ref{eq:eqs22}) as $(1-\mu)(1-\beta_0)+\beta_0 \rho(\bm{B})=1$ and obtain the outbreak threshold as 
\begin{equation}\label{eq:eqs23}
	\beta_c = \frac{\mu}{\rho(\bm{B})+\mu-1},
\end{equation}
where $\rho(\bm{B})$ is the the spectral radius of $\bm{B}$. 
The result suggests that the outbreak threshold is only determined by the network structure and recovery rate, regardless of the intensity of the macro-level social influence.
Specifically, we have $\beta_c = \frac{1}{\rho(\bm{B})}$ if $\mu=1$.

\section{Results}
\label{sec:results}

In this section, our study consists of two parts. 
Firstly, we in turn implement the SAR model on classical networks, including 2D lattices, ER, WS, BA, and SF networks with different heterogeneity and assortativity. 
Subsequently, we refine a physical picture to understand why localization strength determines the magnitude of the tricritical point.

\subsection{Results on Network Models}
\label{subsec:structure}

\begin{figure}[htbp]
	\centering
	\includegraphics[width=0.65\textwidth]{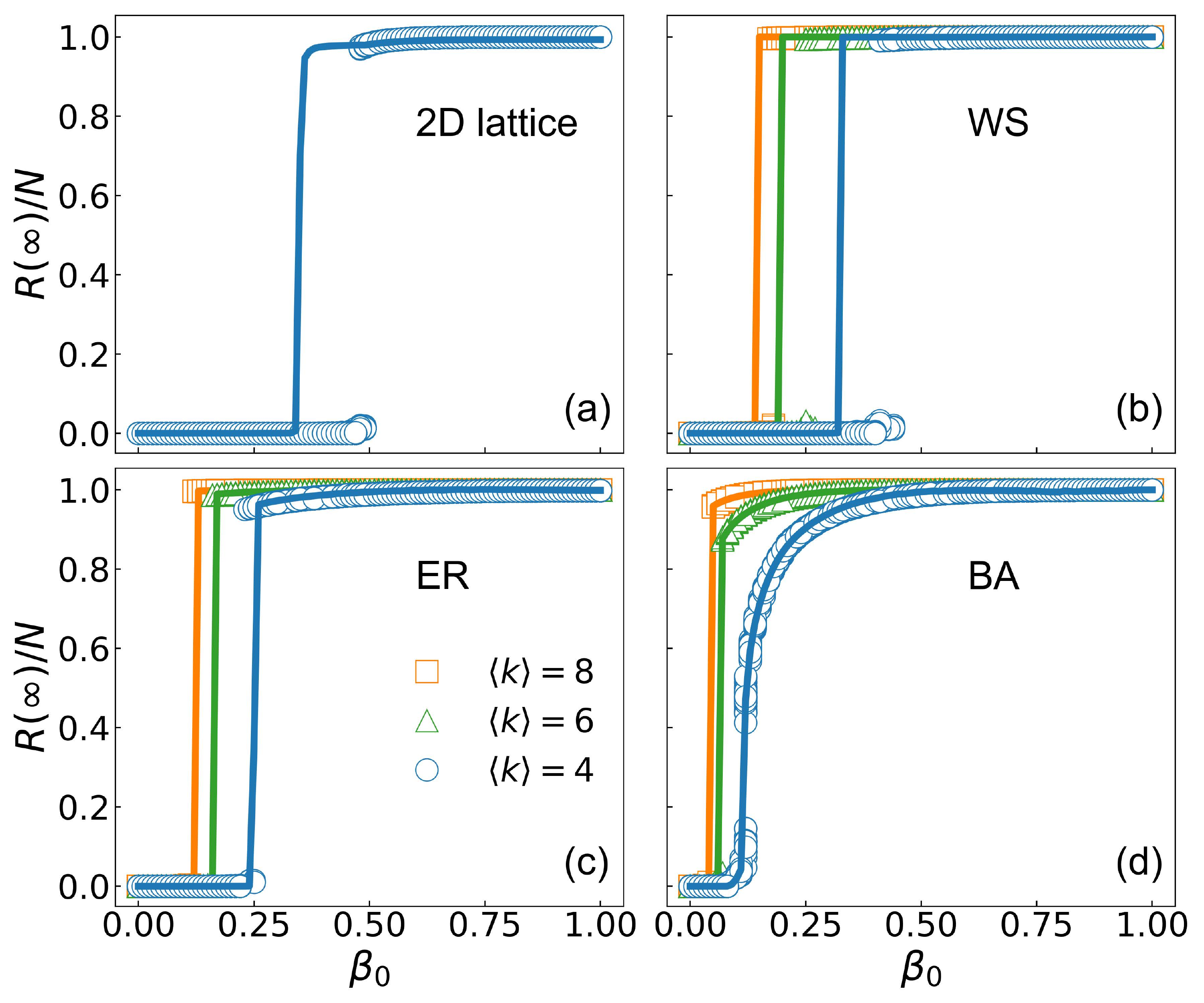}
	\caption{Understanding the effects of network density on spreading dynamics. The density of final adopted population $R(\infty)/N$ is plotted versus intrinsic attractiveness of medium $\beta_{0}$ for four types of classical networks of size $N=10^5$: (a) 2D lattices, (b) WS, (c) ER and (d) BA network. In each subplot, the results with $\langle k \rangle = 4, 6,8$ are plotted, where each marker denotes a single numerical realization of the developed SAR model with $\alpha=1$ and colored solid lines are obtained by the DMP-based approach.}
	\label{fig:Fig1}
\end{figure}

Figure~\ref{fig:Fig1} shows the final adopted populations obtained from the developed SAR model, on classical networks such as 2D lattices, WS, ER, BA, with different mean degrees $\langle k \rangle=4,~6,~8$. 
Definitely differing from the dynamic scenarios formulated by the SIR model, the macro-level social influence can widely induce a discontinuous phase transition. 
More specifically, $R(\infty)/N$ abruptly jumps to $1$ at the threshold on 2D lattices, attributing to that the nodes have the same regular connections such that PFDF can constantly push the contagion process forward. 
However, networks with the same density present smaller jumps with the increase of the degree heterogeneity from WS, ER, to BA [Figs.~\ref{fig:Fig1}(b)-(d)], suggesting that the intensity of feedback is gradually weakened. 
The effect of density on the critical behavior of the system also depends on the heterogeneity of degree, highlighting that both zero- and first-order statistics of networks determine the critical behaviors.
Overall, the results in Fig.~\ref{fig:Fig1} seem to indicate a discontinuous transition as long as there is a macro-level social influence in the process of adoption decision, regardless of the structure details of the networks. 
Furthermore, the DMP-based approach agrees well with numerical simulations except for 2D lattices and WS, because numerous short-loop structures break the assumption of independence between nodal states.

\begin{figure}[htbp]
	\centering
	\includegraphics[width=0.65\textwidth]{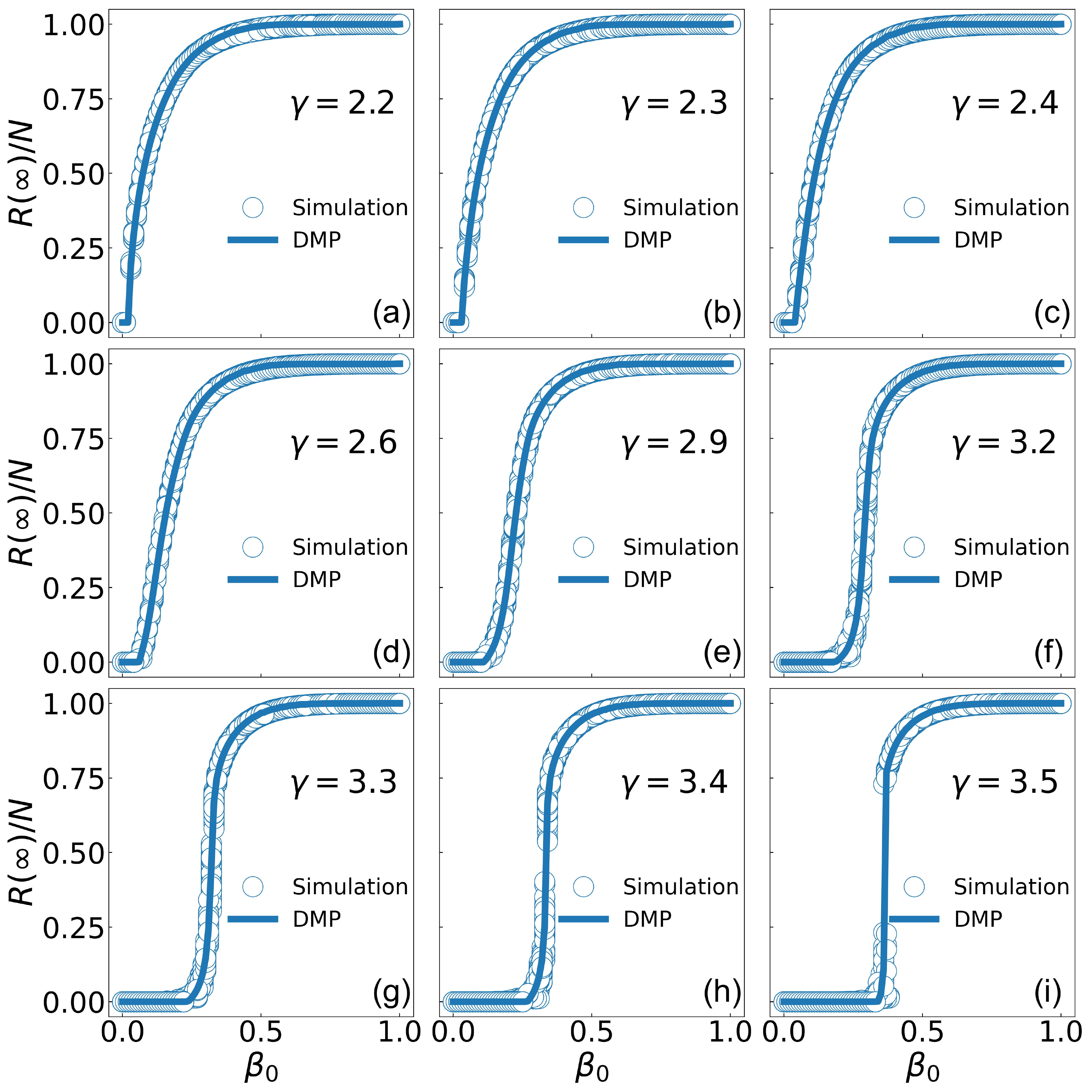}
	\caption{Understanding the effects of network heterogeneity on spreading dynamics. 
		Numerically obtained (blue circles) and theoretically predicted (solid lines) final fractions of the adopted population ($R(\infty)/N$) are plotted versus $\beta_{0}$ for SF networks with different values of $\gamma$. 
		The network size, the parameter $\alpha$, and the numerical simulation are the same as that in Fig.~\ref{fig:Fig1}. 
		For $\gamma=2.4,3.3$, the height of the jump is close to zeros, causing the nature of phase transition is not clear-cut. 
		According to the change from $\gamma=2.2$ to $3.5$, we infer that $\gamma=2.4, 3.3$ is at the boundary between continuity and discontinuity.}
	\label{fig:Fig2}
\end{figure}

More insights are obtained by investigating the effect of network heterogeneity on the nature of phase transition. 
Here, we generate the SF network with different degree exponents $\gamma$ by means of a configuration model. 
More structure details of the SF network are listed in Table~\ref{tab:Tab1} of Appendix~\ref{sec:table}. 
Correspondingly, Fig.~\ref{fig:Fig2} illuminates the complex effects of network heterogeneity in tuning the nature of phase transitions. 
When the network is nearly star-like ($\gamma$ approaching $2$), $R(\infty)/N$ undergoes a discontinuous transition [Figs.~\ref{fig:Fig2}(a)-(c)].
A striking phenomenon then is exhibited, that continuous transitions instead emerge as $\gamma$ increase to the intermediate range of the scale-free regime ($2.6 \lesssim\gamma\lesssim 3.2$) [Figs.~\ref{fig:Fig2}(d)-(f)]. 
However, in the regime of random networks~($\gamma>3.2$), again, the continuous transition is replaced by discontinuous class [Figs.~\ref{fig:Fig2}(g)-(i)]. 
The above phenomena are contrary to our expectation based on the results given by Fig.~\ref{fig:Fig1}, that macro-level social influence is destined for a discontinuous transition; and unlike the conclusions drawn from previous threshold models concerning reinforcement, that increasing network heterogeneity would lead to a continuous transition~\cite{Watts2002}. 
Note the good agreement between numerical and theoretical results, which validates the DMP-based approach. 
It is claimed that the unusual shift of phase transition with heterogeneity is related to the spatial configurations of nodes, naturally, indicating the necessity to further understand the effects of the second-order statistics, i.e. network assortativity. 

\begin{figure*}[tbp]
	\centering
	\includegraphics[width=\textwidth]{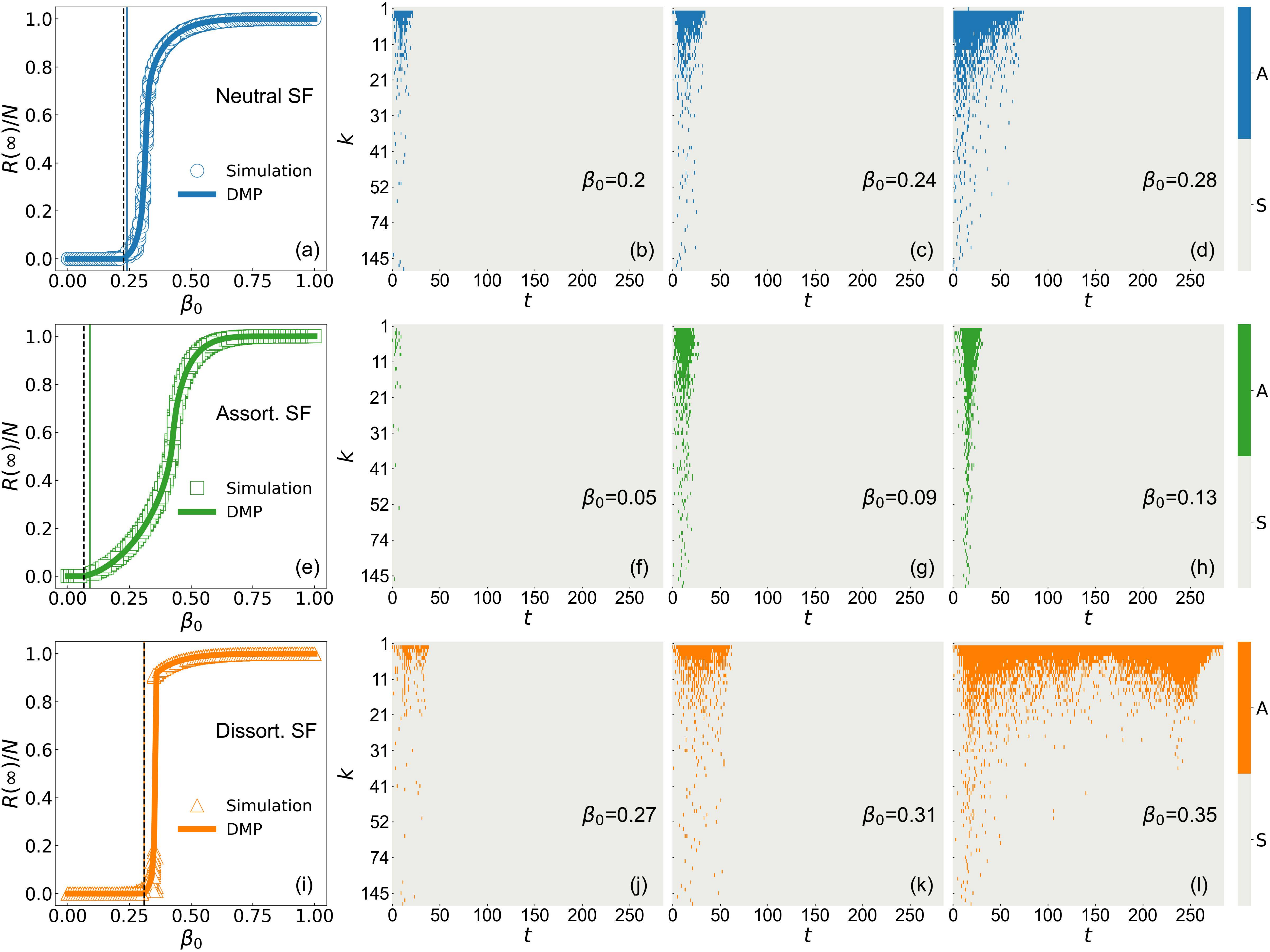}
	\caption{Illustrating the effects of network assortativity on phase transitions. 
		In (a)(e)(i), numerically obtained (markers) and theoretically predicted (solid lines) final fractions of the adopted population ($r(\infty)$) are respectively plotted versus $\beta_{0}$, for neutral, assortative, and disassortative SF networks with $\gamma=3.3$. 
		The assortative~($r=0.170$) and disassortative networks~($r=-0.057$) are generated by the Xulvi-Brunet-Sokolov algorithm, without changing the degrees of nodes.
		The configurations of simulations are the same as that in Fig.\ref{fig:Fig1}. 
		Black vertical lines mark the position of the outbreak thresholds estimated in terms of the susceptibility $\chi=[\langle r(\infty)^2 \rangle - \langle r(\infty) \rangle^2]/\langle r(\infty) \rangle$), and colored vertical lines the theoretical thresholds given by Eq.~(\ref{eq:eqs23}). Correspondingly, the evolution of adopted nodes for different degree classes around the outbreak threshold is illustrated for the neutral network (b)-(d), assortative network (f)-(h), and disassortative network (j)-(l).}
	\label{fig:Fig3}
\end{figure*}

The effects of network assortativity on the nature of phase transition are elucidated by varying the assortativity of SF networks, as shown in Fig.~\ref{fig:Fig3}. 
Here, we employ the Xulvi-Brunet and Sokolov algorithm to generate either assortative or disassortative networks, by preserving the same degree sequence as neutral SF networks~\cite{Xulvi2004}. 
Interestingly, in spite of following the identical degree sequence, increasing assortativity alone can change the transition from being abrupt to continuous. 
Figure~\ref{fig:Fig3}(b)-(d),(f)-(h),(j)-(l) provide a visual picture of the results in Figs.~\ref{fig:Fig3}~(a),~(e) and (i), which give a clear understanding.  
In the assortative network, the adoption seldom occurs in nodes with small degrees, it tends to localize near the hubs under the macro-level social influence [Figs.~\ref{fig:Fig3}(f)-(h)].  
The degree correlation coefficient of the neutral network is slightly positive, hubs do not have sufficient feedback to drive non-hubs [Figs.~\ref{fig:Fig3}(b)-(d)], exhibiting a continuous transition. 
Figure~\ref{fig:Fig3}(j)-(l) document a much more sustainable process, i.e., an avalanche-like outbreak happens as the degree correlation coefficient $r$ decreases to be negative.
Notice that the threshold becomes larger with decreasing assortativity, in accordance with the interpretation put forward in Ref.~\cite{new2002}. 

\begin{figure*}[htbp]
	\centering
	\includegraphics[width=\textwidth]{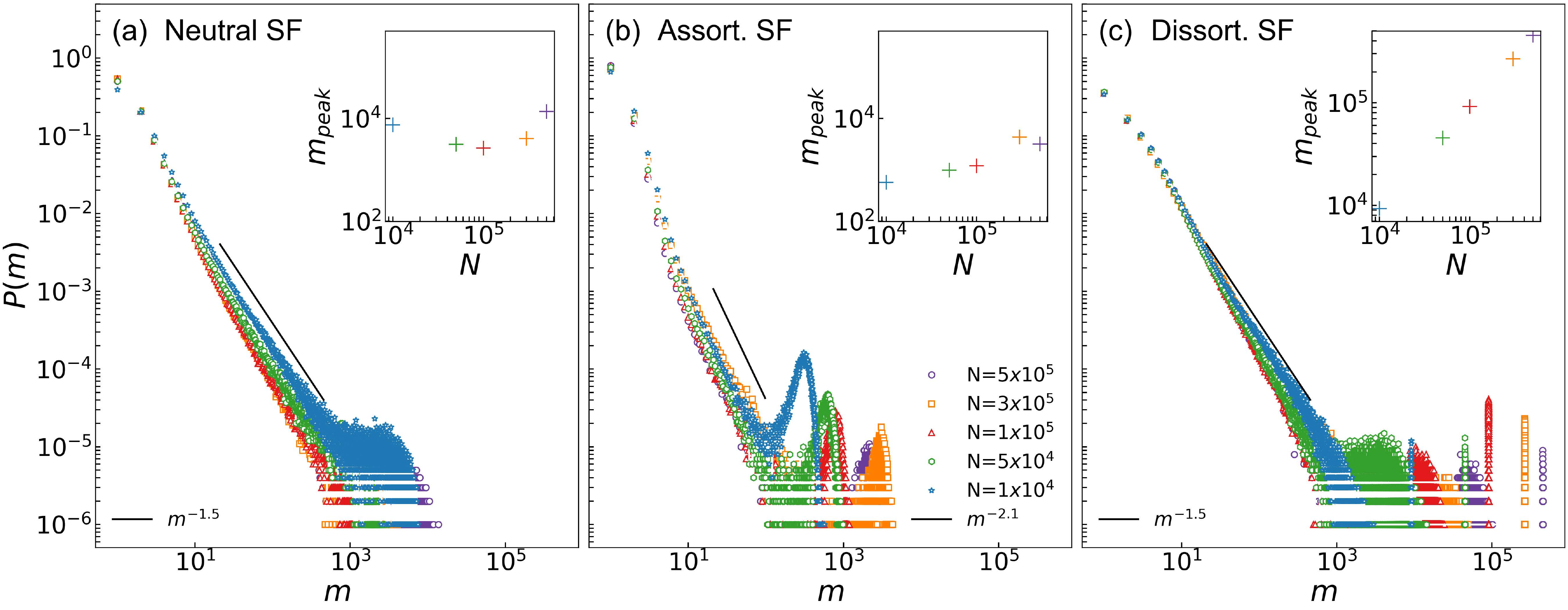}
	\caption{The nature of phase transition on the SF networks of different assortativity: (a) neutral network, (b) assortative network, and (c) disassortative network. 
		Mass distribution of the final adopted population at the outbreak threshold is obtained from $10^6$ realizations. 
		The size of the employed networks ranges from $N=10^4$ to $5\times10^5$. $\gamma=3.3$ and $\alpha=1$. 
		The solid straight lines represent the decay, where the exponent is approximated through the maximum likelihood method. 
		The peaks on the right side, separated clearly from the left power-law distribution, corresponds to giant adopted clusters. 
		The emergence of a giant adopted cluster and linear scaling relationship $m_{peak}\sim N$ indicates a discontinuous transition, and a continuous transition otherwise. 
		$m_{peak}$ denotes the size of the giant adopted cluster.}
	\label{fig:Fig4}
\end{figure*}

Still, the DMP-based approach provides accurate predictions, with respect to not only the final adopted fractions of the population but also the position of the threshold [Fig.~\ref{fig:Fig3}(a)(e)(i)]. 
We assume that the threshold does not change with the macro-level social influence, which is proved to be rational by the results illustrated in Fig.~\ref{fig:Sfig2}.
The threshold does not depend on $\alpha$ and is equal to the threshold of the classical SIR. 
One clearly observed that the theoretical lines and numerical markers highly coincide with each other, regardless of assortativity~(See Appendix~\ref{apec:evolution} for more details). 
Moreover, the conclusion about the nature of the transitions in Fig.~\ref{fig:Fig3} is strongly supported by the simulation results illustrated in 
Fig.~\ref{fig:Fig4}. 
As the networks change from assortative to disassortative, the nature of the transition passes from continuous to discontinuous, along with a larger threshold.

\subsection{A comprehensive understanding of the joint effects}

\label{subsec:picture}

\begin{figure*}[htbp]
	\centering
	\includegraphics[width=0.65\textwidth]{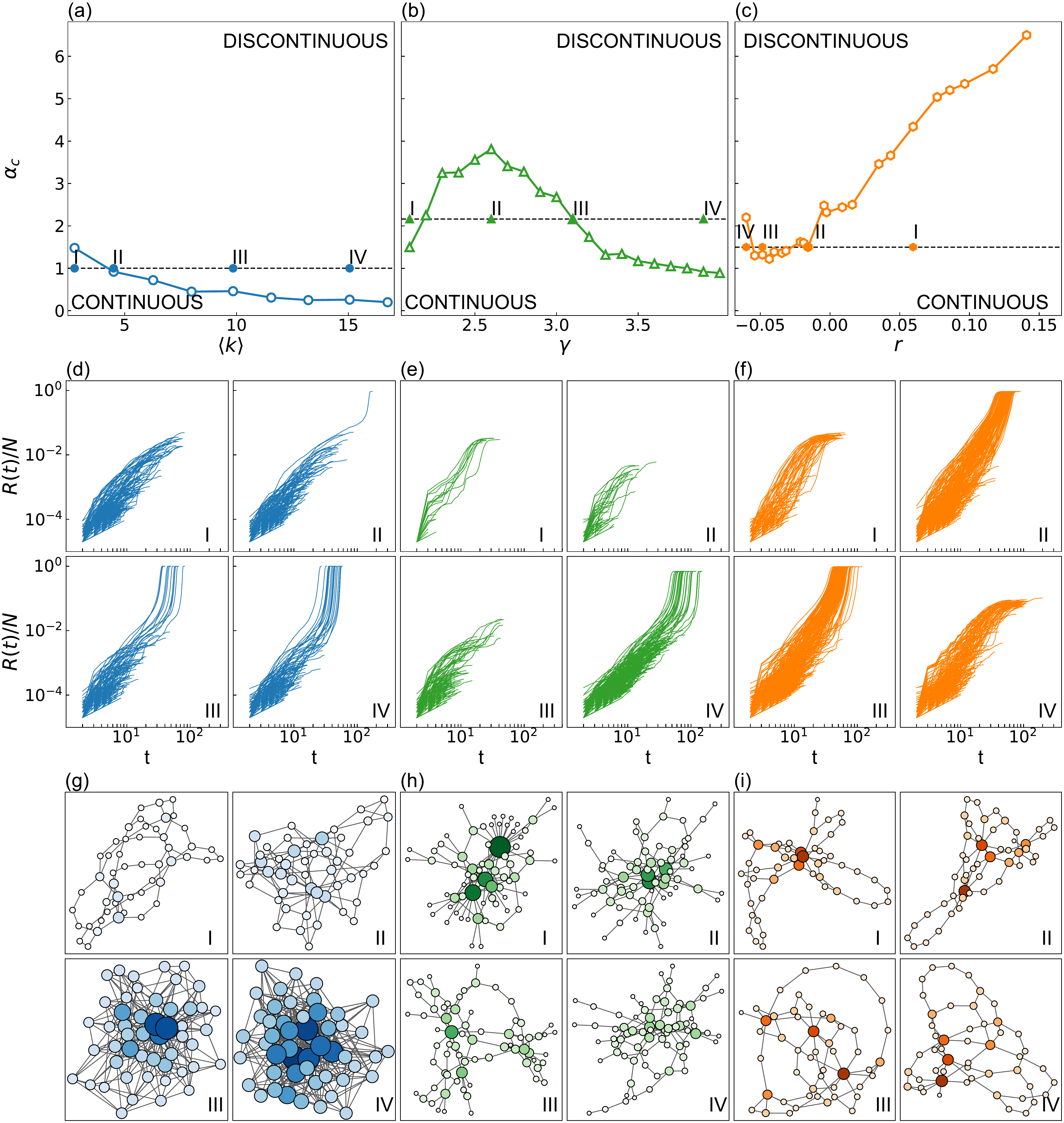}
	\caption{Understanding the joint effects of network structures and macro-level social influence. 
		(a)-(c) The tricritical point $\alpha_c$ as function of mean degree $\langle k \rangle$, degree exponent $\gamma$, and degree correlation coefficient $r$, respectively. 
		$\alpha_c$ is identified through the patterns of mass distribution of $r(\infty)$ at the theoretical $\beta_c$, as shown in Fig.~\ref{fig:Fig4}. 
		The size of networks is $N = 10^5$. 
		In (a), the SF networks with different degrees are generated by the configuration model, by increasing the minimum degree $k_{min}$ from $2$ to $10$. 
		The positions of the four labeled parameter points are $\langle k \rangle = 2.75$, $\langle k \rangle = 4.50$, $\langle k \rangle = 9.84$ and $\langle k \rangle = 15.06$, with $\alpha=1.0$. 
		In (b), we generate the SF network with different $\gamma$ by the Goh-Kahng-Kim algorithm put forward in Ref.~\cite{Goh2001}. 
		The positions of the four labeled parameter points are $\gamma=2.1$, $\gamma=2.6$, $\gamma=3.1$ and $\gamma=3.9$, with $\alpha=2.16$. 
		In (c), the networks with different $r$ are generated by the Xalvi-Brunet-Sokolov algorithm, starting with a neutral scale-free network with $\gamma=3.3$ and $\langle k\rangle =2.75$. 
		The positions of the four labeled parameter points are $r=-0.060$, $r=-0.048$, $r =-0.016$, and $r=0.059$, with $\alpha=1.0$. 
		For each subplot in (a)-(c), the dynamic evolution of the order parameter $R(t)/N$ at theoretical $\beta_c$ are illustrated in (d)-(f), corresponding to the labeled points, i.e. \uppercase\expandafter{\romannumeral1}, \uppercase\expandafter{\romannumeral2}, \uppercase\expandafter{\romannumeral3}, \uppercase\expandafter{\romannumeral4} in (a)-(c). 
		The simulation results are from 2000 realizations, where each realization represents a curve. 
		In (g)-(i), the structural patterns of size $N^{'}=50$ give a more intuitive picture, without leaving the key structural properties of the employed networks in (d)-(f). 
		Darker and larger nodes indicate more connections they have.}
	\label{fig:Fig5}
\end{figure*}

The aforementioned results for classic networks highlight three structure terms that can regulate the macro-level social influence on phase transitions: density (or mean degree), heterogeneity, and assortativity, as 0-order, 1-order, and 2-order statistics of the network. 
A continuous transition occurs if relatively heterogeneous networks are assortative and sparse enough.
However, a clear physical picture of the joint effects of network structure and macro-level social influence is still lacking. 
In this section, we will give a comprehensive understanding of the dependence of the nature of phase transition on strength of macro-level social influence induced by structural patterns of SF networks.

\begin{figure*}[htbp]
	\centering
	\includegraphics[width=\textwidth]{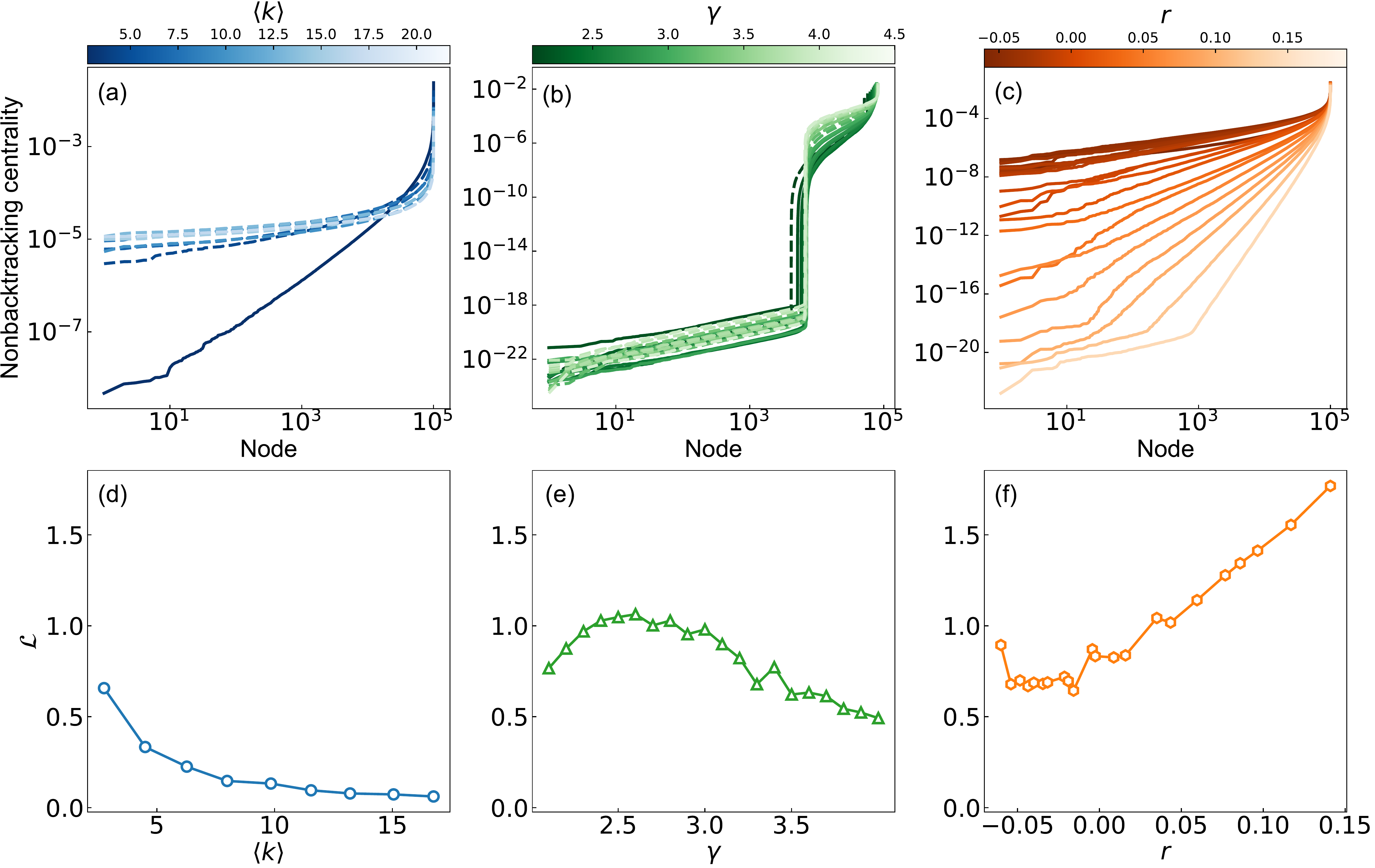}
	\caption{Localization effect on the networks corresponding to Fig.~5. 
		(a)-(c) Non-backtracking centrality $f_i(\Lambda_1)$ of each node in the network is exhibited in ascending order. 
		The continuous transition is plotted with solid lines if $\alpha_c > \alpha$, otherwise dashed. 
		(a) $\alpha=1$. (b) $\alpha=2.16$. (c) $\alpha=1$.  
		(d)-(f) Localization strength~($\mathcal{L}$) versus structural quantities. $\mathcal{L} =\frac{1}{\langle k\rangle}(\frac{1}{N}\sum_{i=1}^N(f_i(\Lambda_1)-\langle f_i(\Lambda_1) \rangle)^2)^{1/2}$ where $f_i(\Lambda_1)$ denotes the non-backtracking centrality of node. A larger $\mathcal{L}$ corresponds to a stronger localization effect in networks.
	}
	\label{fig:Fig6}
\end{figure*}

We vary individual structural parameters to generate the networks, with other structure variables remaining fixed as much as possible. 
Figure~\ref{fig:Fig5} focuses on how the critical strength of macro-level social influence $\alpha_c$, beyond which the transition changes from continuous to discontinuous, behaves with network density (mean degree), heterogeneity, and assortativity, respectively. 
Figure~\ref{fig:Fig5}(a) shows an overall decrease of $\alpha_c$ with mean degree $\langle k\rangle$, but a more slow descent as the networks become rather dense. 
This is in agreement with what is found in Fig.~\ref{fig:Fig1}, that the denser the network is [Fig.~\ref{fig:Fig5}(g)], the more easily macro-level social influence can facilitate an avalanche-like contagion [Fig.~\ref{fig:Fig5}(d)]. 
Sparse networks have many smaller-degree nodes and long-loops~[Fig.~\ref{fig:Fig5}(g\uppercase\expandafter{\romannumeral1})], contributing to a weaker PFDF and thus a continuous transition [Fig.~\ref{fig:Fig5}(d\uppercase\expandafter{\romannumeral1})]. 
A distinctive behavior observed in the regions of discontinuous transitions is an additional explosive growth [Fig.~\ref{fig:Fig5}(d\uppercase\expandafter{\romannumeral2})(d\uppercase\expandafter{\romannumeral3})(d\uppercase\expandafter{\romannumeral4})], since the contagion can reach the hubs and becomes avalanche-like through sufficient available channels to those small-degree nodes [Fig.~\ref{fig:Fig5}(g\uppercase\expandafter{\romannumeral2})(g\uppercase\expandafter{\romannumeral3})(g\uppercase\expandafter{\romannumeral4})].  

Nevertheless nonlinear dependence of $\alpha_c$ on degree exponent $\gamma$ and degree correlation coefficient $r$ can be correspondingly observed in Figs.~\ref{fig:Fig5}(b) and (c). 
More in detail, in Fig.~\ref{fig:Fig5}(b), $\alpha_c$ firstly increases to about $3.8$ in the intermediate scale-free regime and then keep decreasing to a flat region in the random network regime. 
It is evident that a discontinuous transition is more likely to occur if connections are either extremely heterogeneous or homogeneous, in line with the conclusions drawn from Fig.~\ref{fig:Fig2}. 
Discontinuous transition can be found at the parameter points \uppercase\expandafter{\romannumeral1} and \uppercase\expandafter{\romannumeral4}, however, the mechanisms lying behind are fundamentally different.
With small $\gamma$, star-like SF network owns more numerous small-degree nodes, as well as super hubs, which together bring about larger mean degree [Table.~\ref{tab:Tab1} and Fig.~\ref{fig:Fig5}(h\uppercase\expandafter{\romannumeral1})]. 
Due to a considerable number of close contacts and shorter length of the path, the super hubs can readily push the contagion to penetratingly reach the patterns of small-degree nodes, to some extent which weakens the macro-level social influence [see Fig.~\ref{fig:Fig5}(e\uppercase\expandafter{\romannumeral1}) and Fig.~\ref{fig:Fig2}(a) for jumps of smaller size]. 
%Most nodes would get adopted once the super hubs are trying to transform the medium. 
In cases of large $\gamma$ (e.g. parameter point \uppercase\expandafter{\romannumeral4}), hubs continue to be present, but small and less numerous, the difference in the degree of individuals is thus insignificant, as shown in Fig.~\ref{fig:Fig5}(h\uppercase\expandafter{\romannumeral4}). 
There are abundantly available transmission channels among hubs and non-hubs for the contagion, rather than being simply impeded by hubs. 
Hence,  the macro-level social influence persistently reinforces individual adoption probability, more easily leading to an avalanche-like outbreak~[Fig.~\ref{fig:Fig5}(e\uppercase\expandafter{\romannumeral4})].
The picture is obviously distinct when the networks are in an intermediate range of $\gamma$ (e.g. parameter point \uppercase\expandafter{\romannumeral2}), where hubs are numerous and cluster together, such that they have insufficient connections to make the contagion immediately stretch over the "periphery architecture" consisting of small-degree nodes~[Fig.~\ref{fig:Fig5}(h\uppercase\expandafter{\romannumeral2})]. 
This is confirmed by the temporal evolution patterns that the growth rate is relatively low, along with a short duration [Fig.~\ref{fig:Fig5}(e\uppercase\expandafter{\romannumeral2})]. 
It is consequently easy to have a continuous transition.
In addition, parameter point \uppercase\expandafter{\romannumeral3} is at the boundary between continuous and discontinuous transition, at which the temporal evolution pattern has no avalanche [Fig.~\ref{fig:Fig5}(e\uppercase\expandafter{\romannumeral3})].

Figure~\ref{fig:Fig5}(c) shows that increasing $r$ has a monotonous growth of $\alpha_c$, except for a slight drop when the network is disassortative. 
The generated assortative network contains a tight core of nearly all hubs through which numerous long loops pass, surrounded by sparse periphery architectures of small-degree nodes [Fig.~\ref{fig:Fig5}(i\uppercase\expandafter{\romannumeral1})]. 
Therefore, it is not easy for the contagion to persist along these long loops for a long time [Figs.~\ref{fig:Fig3}(f)-(h) and Fig.~\ref{fig:Fig5}(f\uppercase\expandafter{\romannumeral1})], especially after that it stretches into the periphery architectures far away from the hubs. 
As the network approaches a disassortative state, hubs are increasingly partitioned by small-degree nodes [Fig.~\ref{fig:Fig5}(i\uppercase\expandafter{\romannumeral2})(i\uppercase\expandafter{\romannumeral3})], and hubs have more chances to connect with non-hubs. 
In such cases, the hubs play the role of "Destination Charger" with the assistance of the  macro-level social influence, which can drive the medium to diffuse through surrounding circles of non-hubs to reach other hubs, triggering further contagion. 
The above process repeats constantly and finally gives rise to an avalanche-like outbreak [Fig.~\ref{fig:Fig5}(f\uppercase\expandafter{\romannumeral2})(f\uppercase\expandafter{\romannumeral3})]. 
However, a more drastic avalanche cannot be expected as $r$ decreases to be rather negative [Fig.~\ref{fig:Fig5}(c\uppercase\expandafter{\romannumeral4})], attributing to a long distance between the hubs and that the non-hubs with a larger degree are moving away. 
There are thus more long loops along which the contagion frequently fails to persist [Fig.~\ref{fig:Fig5}(i\uppercase\expandafter{\romannumeral4})]. 

We finally uncover a sufficient condition for a continuous transition: simultaneously, being sufficiently sparse, heterogeneous, either assortative or rather disassortative; otherwise, the transition is instead discontinuous. 
However, one needs to know what mechanism ultimately dominates the nature of phase transition in the macro-level social influence, whose effects can also be regulated by network density, heterogeneity, or assortativity. 
Figure~\ref{fig:Fig6}(a)-(c) plot the non-backtracking centrality of nodes in ascending order, for the networks employed in Fig.~\ref{fig:Fig5}(a)-(c), which provides key hints on the issue.
Combining with results in Fig.~\ref{fig:Fig6}, we find that, if the divergence of the non-backtracking centrality is sufficiently great, i.e. more apparent height difference between the plain and the platform or peak, the transition is continuous; and discontinuous otherwise. 
The nodes lying in the plain belong to periphery structure patterns, in contrast to those hubs constructing a tight core located in the platforms or peaks. 
While the core-periphery structure indicates strong localized effects, which is the mechanism we are seeking for. 
An avalanche happens on condition that the contagion is not localized in the core pattern, and persistently preserved by the "Destination Chargers"-hubs to stretch over the periphery areas. 
In other words, more likely, we will have a continuous transition if the localization effects become pronounced, meaning that there is a tight core pattern to capture a considerable fraction of connections, leaving fewer connections to the periphery structure (this is why the whole network must be sparse). 
Accordingly, we use the metric $\mathcal{L}$ to quantify the localization strength of networks~\cite{Xue2022}.  
Figure~\ref{fig:Fig6}(d)-(f) show that the changing trend of $\mathcal{L}$ is qualitatively consistent with $\alpha_c$ by varying the network density, heterogeneity, and assortativity, revealing that the magnitude of the tricritical point is governed by the localization strength of network structure.
Therefore, it is easier to preserve a continuous transition if $\mathcal{L}$ in the network is larger, which provides a basis to evaluate the phase transition for real networks. 
While the measures reducing the localization effect of the network, such as increasing the mean degree and making the network less heterogeneous or assortative, can change the transition from continuous to discontinuous.

\section{Discussion}
\label{sec:conclusion}
In this paper, we have performed analysis of the innovation diffusion model incorporating both the micro and macro level social influence on classical synthetic networks. 
In addition, we have applied the DMP method to track the final fractions of recovered population, and to derive an expression for the outbreak threshold which does not depend the intensity of the macro social influence, in agreement with simulations. 
With a given intensity of macro social influence of $\alpha=1$, a discontinuous transition is still exhibited on classical networks such as 2D lattices, WS, ER, and BA networks, regardless of the structure details. 
Moreover, network heterogeneity is the premise to unfold the effects of network density in facilitating a discontinuous phase transition. 
Then, an interesting and striking point regards what happens for SF networks, i.e., the transition is continuous when the network is in the intermediate range of scale-free regime ($2.6 \lesssim\gamma\lesssim 3.2$). 
We further tune the assortativity of the SF network with $\gamma=3.3$ by means of the Xulvi-Brunet and Sokolov algorithm, uncovering that whether continuous transition can be observed depends on that positive feedback capable of promoting the contagion tends to be localized near those hubs, which requires an assortative structure. 

In the second part of this paper, we provided a comprehensive understanding of the dependence of tricritical point $\alpha_c$ on three structure quantities: network density, heterogeneity, and assortativity, giving a clear physical picture of their joint effects on the localization strength. 
In more detail, we find that, apart from a weak intensity of macro-level social influence, sparsely available connections, and relatively heterogeneous distribution, either assortative or extremely disassortative configurations are simultaneously indispensable for a continuous transition. 
The core-periphery structure is thus required. 
Otherwise, a discontinuous transition is instead more easily achieved. 
The hubs persistently preserve the contagion to have an avalanche, when there are sufficient available channels, the network is either star-like or homogeneous, or there are appropriate intervals of non-hubs between the hubs i.e., being appropriately assortative. 
We further report a strong confirmation of the association between the tricritical point and localization effects. The issue can be clarified by plotting the spectrum of non-backtracking centrality of networks, and the localization strength measured by the metric developed in Ref.~\cite{Xue2022}. 
More specifically, core-periphery structures yield a strong localization effect, and thus a continuous transition. 
Our analysis highlights that core-periphery structure, being dependent on network density, heterogeneity, and assortativity, is the structure pattern that essentially governs localization effects, and the magnitude of the tricritical point in presence of macro-level social influence.

We conclude our study by providing two general remarks.
%First, our study identifies a novel dynamic mechanism to achieve a discontinuous phase transition, that is macro-level social influence, besides reinforcement effects~\cite{Centola2010} and mutual cooperation between multiple diffusion process~\cite{Cai2015}. 
The macro-level social influence turns out to be a non-negligible ingredient in innovation propagation. 
It essentially acts on the diffusion process with a form of positive feedback and is thus capable of inducing a discontinuous transition.
However, based on our results, we can say that strong positive feedback is not a sufficient condition for the occurrence of discontinuous phase transition. It is also up to localization strength governed by core-periphery structure.
Secondly, our study suggests that the positive feedback between the macro and micro scales of complex systems is proved to be a useful tool to underline the effects of the mesoscopic structure patterns such as core-periphery structure on the local micro dynamics. It guides future research direction. 

It should be noted that the network models employed in this manuscript cannot capture all structural characteristics of natural or man-made networks, especially higher-order statistics such as high clustering or community structures found commonly in real-world networks. These factors are  potentially related to the change of core-periphery structure~\cite{Polanco2023}. Therefore, further attempts to explore the relationship of them are still required in the future.
In addition, the investigation of multiple feedback in general complex contagion dynamics remains a very interesting avenue for future research activity. 
We also hope to be able to extend our study to different contexts, as well as to more diverse complex networks such as metapopulation networks~\cite{Wang2018,Gao2022}, temporal networks~\cite{Onaga2017}, multilayer networks~\cite{Soriano2018,DEARRUDA20181}, hypergraphy~\cite{Chowdhary2021,ferraz2021} and those have hierarchical core-periphery structures~\cite{Polanco2023}. 

\section*{Acknowledgments}
This work was supported by the Key Program of the National Natural Science Foundation of China (Grant No.~71731002), and by Guangdong Basic and Applied Basic Research Foundation (Grant No.~2021A1515011975).

\appendix
\section{Statistical characteristic of SF networks} 
\label{sec:table}
\begin{table}[htb]
	\centering
	\caption{The basic statistical characteristic of SF networks generated by configuration model: degree exponent $\gamma$, the number of nodes $N$, the number of edges $L$, mean degree $\langle k \rangle$, assortativity coefficient $r$, where the minimum and maximum degree are respectively set as $k_{min}=1$ and $k_{max}\sim \sqrt{N}$ to avoid structural disassortativity.}
	\begin{tabular}{@{}rrrrr@{}}
		\toprule
		\multicolumn{1}{c}{$\gamma$} & \multicolumn{1}{c}{$N$} & \multicolumn{1}{c}{$L$}& \multicolumn{1}{c}{$\langle k \rangle$} & \multicolumn{1}{c}{$r$} \\ \midrule
		2.2 & 100000 & 296511 & 5.9302 & -0.0049 \\
		2.3 & 100000 & 262034 & 5.2407 & -0.0047 \\
		2.4 & 100000 & 235218 & 4.7044 & -0.0015 \\
		2.6 & 100000 & 197329 & 3.9466 & -0.0012 \\
		2.9 & 100000 & 162283 & 3.2457 & -0.0046 \\
		3.2 & 100000 & 143425 & 2.8685 & 0.0002 \\
		3.3 & 100000 & 137872 & 2.7574 & 0.0044 \\
		3.4 & 100000 & 135253 & 2.7051 & 0.0006 \\
		3.5 & 100000 & 131266 & 2.6253 & -0.0033 \\  \bottomrule
	\end{tabular}
	\label{tab:Tab1}
\end{table}

\section{Outbreak threshold}
\label{apec:evolution}

\begin{figure}[htbp]
	\centering
	\includegraphics[width=0.5\textwidth]{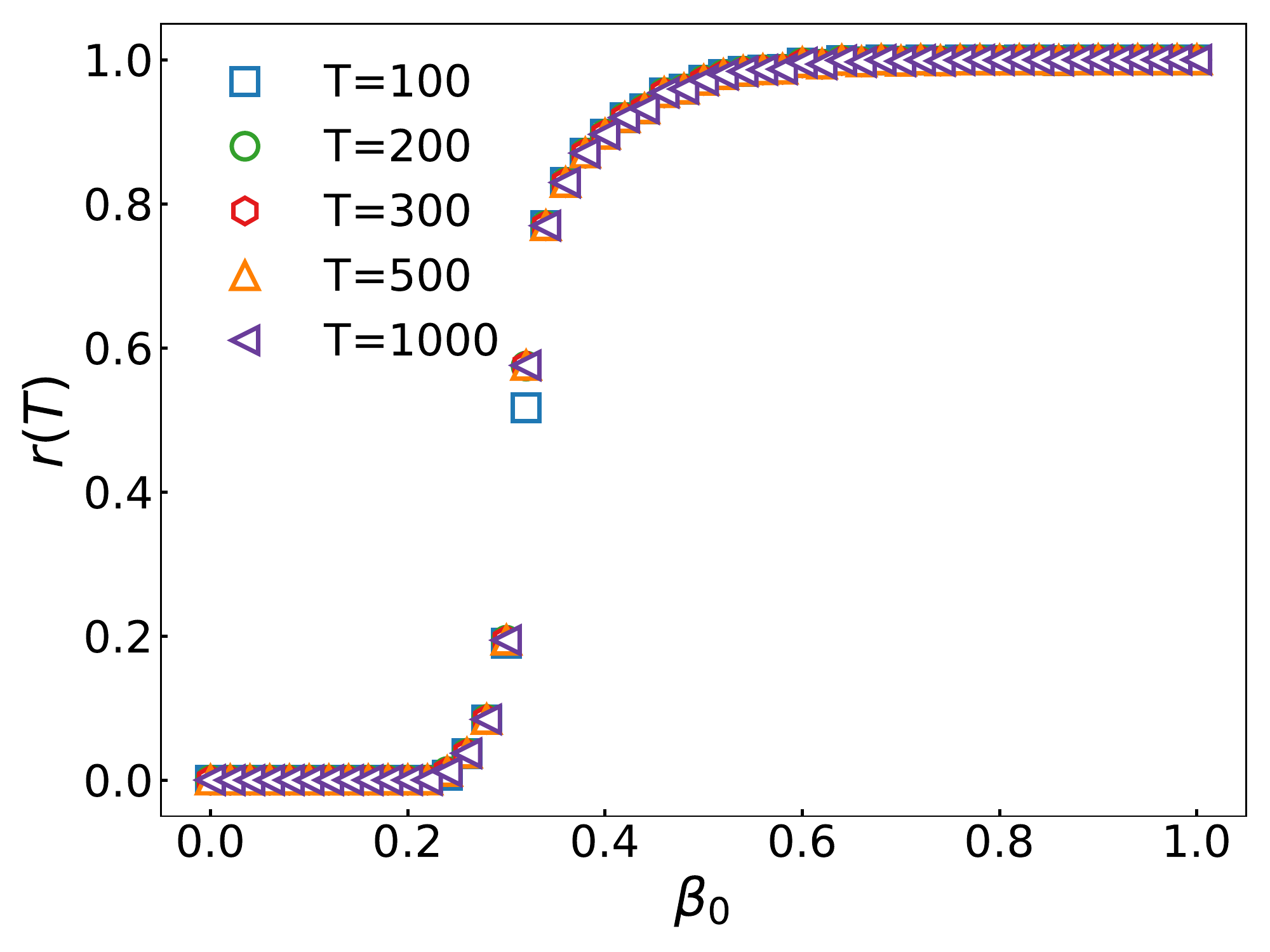}
	\caption{The convergence of DMP. 
		The density of the final adopted population at time $T$ is plotted as a function of the intrinsic attractiveness $\beta_0$. 
		The result is obtained on the neutral scale-free networks ($\gamma=3.3$) by averaging $20$ different initial conditions. }
	\label{fig:Sfig1}
\end{figure}

The density of the final adopted population is plotted as a function of $\beta_0$ for different time steps [Fig.\ref{fig:Sfig1}]. 
For T $\geq$ 200, all points agree well with each other, suggesting that the result obtained through DMP-based approach has converged. 
Therefore, we obtain $r(\infty$) for $T=300$ in this paper. 

\begin{figure*}[htbp]
	\centering
	\includegraphics[width=\textwidth]{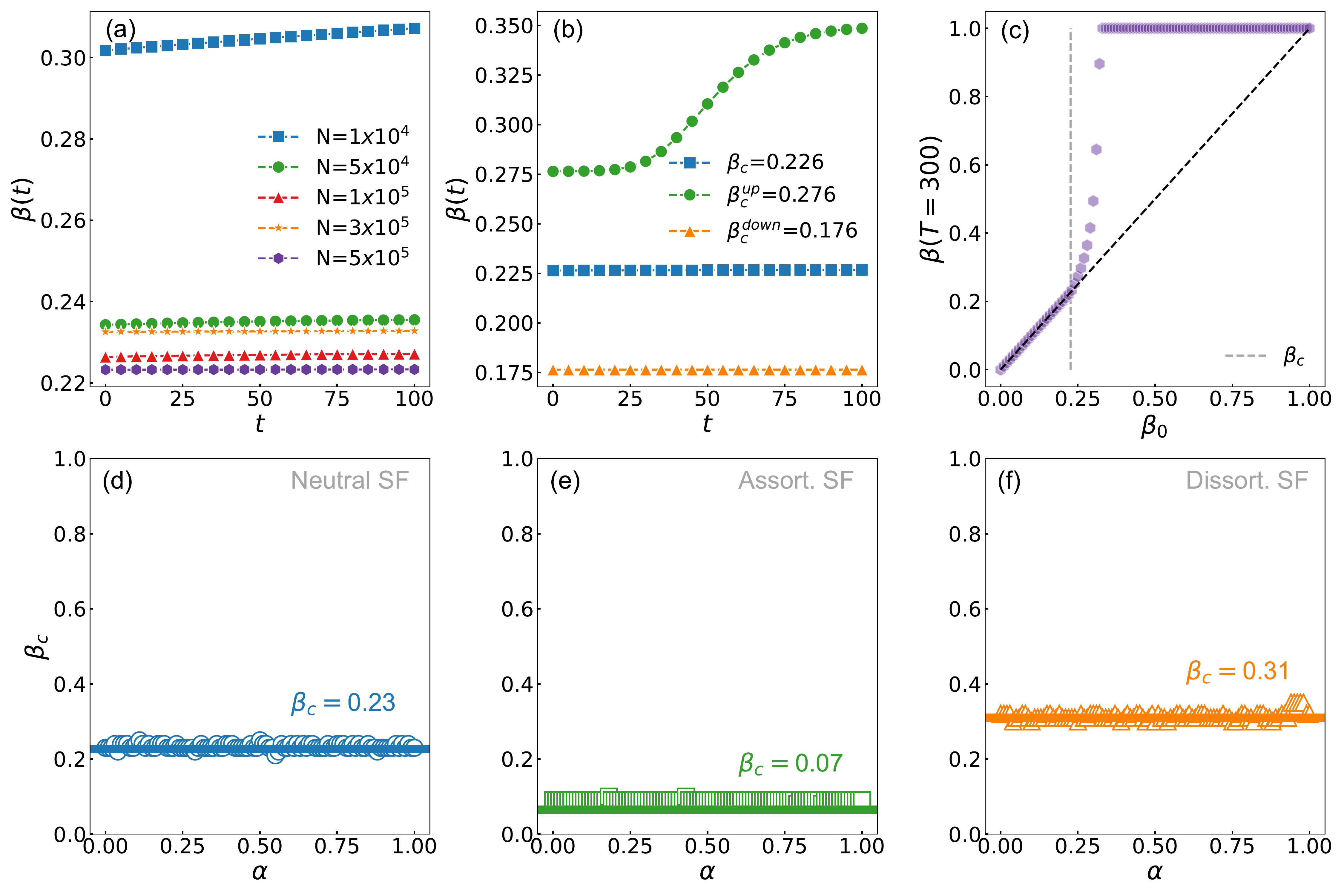}
	\caption{The outbreak threshold derived from the DMP-based approach and numerical simulation. 
		(a) Temporal evolution of $\beta(t)$ for the network of different sizes~($N$) at theoretical outbreak threshold $\beta_c$. 
		(b) Temporal evolution of $\beta(t)$ for different intrinsic attractiveness ($\beta_0=\beta_c$, $\beta_c^{up}$, $\beta_c^{down}$). 
		(c) The adoption probability at $T=300$ is plotted as a function of $\beta_0$. 
		The gray line denotes $\beta_0 =\beta_c$, and a black line with a slope of 1 is used to guide the eye such that one observes clearly $\beta(t=300) = \beta_0$ when $\beta_0 \leq \beta_c$. 
		(d)-(f) The outbreak thresholds under different $\alpha$ are identified by the DMP-based approach~(solid line) and the susceptibility~(markers).         
		The employed SF networks have different assortativity: (d) neutral, (e) assortative, and (f) disassortative networks. 
		The network size is $N=10^5$, and $\gamma=3.3$.
	}
	\label{fig:Sfig2}
\end{figure*}

The phase transition derived by the DMP-based approach does not change with the macro-level social influence as the system approaches the thermodynamic limit~($N\rightarrow \infty$).
Here, we verify that the argument still holds in a large-scale system by the temporal evolution of adoption probability. 
Fig.~\ref{fig:Sfig2}(a) shows that $\beta(t)$ does not change with the evolution of $t$ when the size of the network reaches the $10^5$, suggesting that the macro-level social influence does not work at theoretical $\beta_c$.
However, $\beta(t)$ changes dramatically with $t$ once a small perturbation occurs above $\beta_c$~[Fig.~\ref{fig:Sfig2}(b)].
A clear result is shown in Fig.~\ref{fig:Sfig2}(c), $\beta(t)$ starts to deviate from the diagonal when $\beta_0 > \beta_c$.
In addition, outbreak thresholds identified by susceptibility on networks with different degree correlation coefficients do not depend on $\alpha$~[Fig.~\ref{fig:Sfig2}(d)-(f)], further supporting the conclusion that the position of phase transition does not change with the macro-level social influence.

\section{The effect of multiple seeds}
\label{apec:evolution}
\begin{figure*}[!h]
\centering
\includegraphics[width=\linewidth]{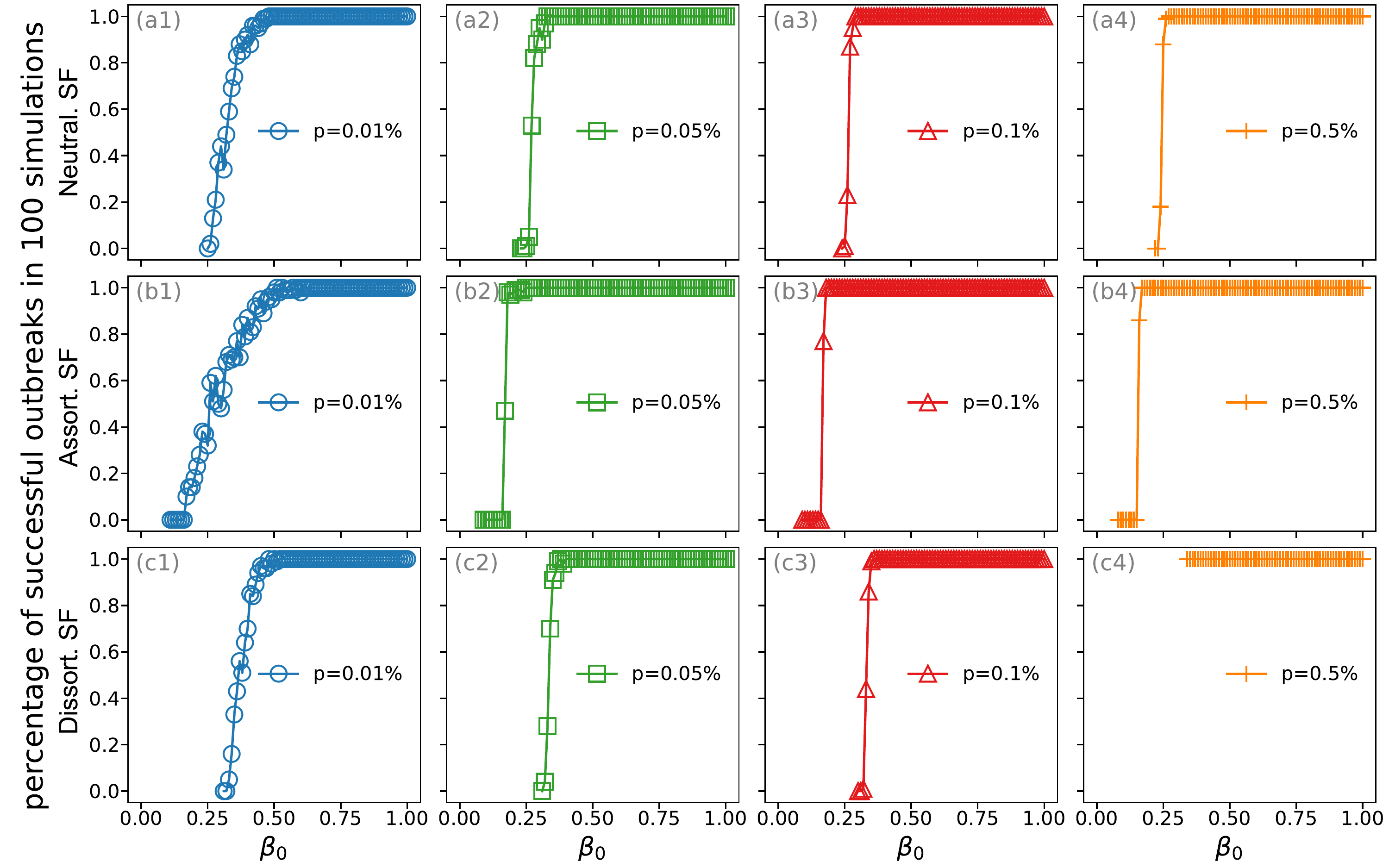}
\caption{
The effect of seed number on the proportion of successful outbreaks (i.e., $R(\infty)/N>5\%$).
We show the results on (a1)-(a4) neutral, (b1)-(b4) assortative and (c1)-(c4) disassortative SF networks, which are the same as those used in Fig.~\ref{fig:Fig3} in the main text. The value of $p$ indicates the percentage of seed nodes in each case.
The results are obtained from $100$ independent realizations.
The zero-value markers are excluded for clarity. 
}
\label{fig:Sfig3}
\end{figure*}

\begin{figure*}[!h]
\centering
\includegraphics[width=\linewidth]{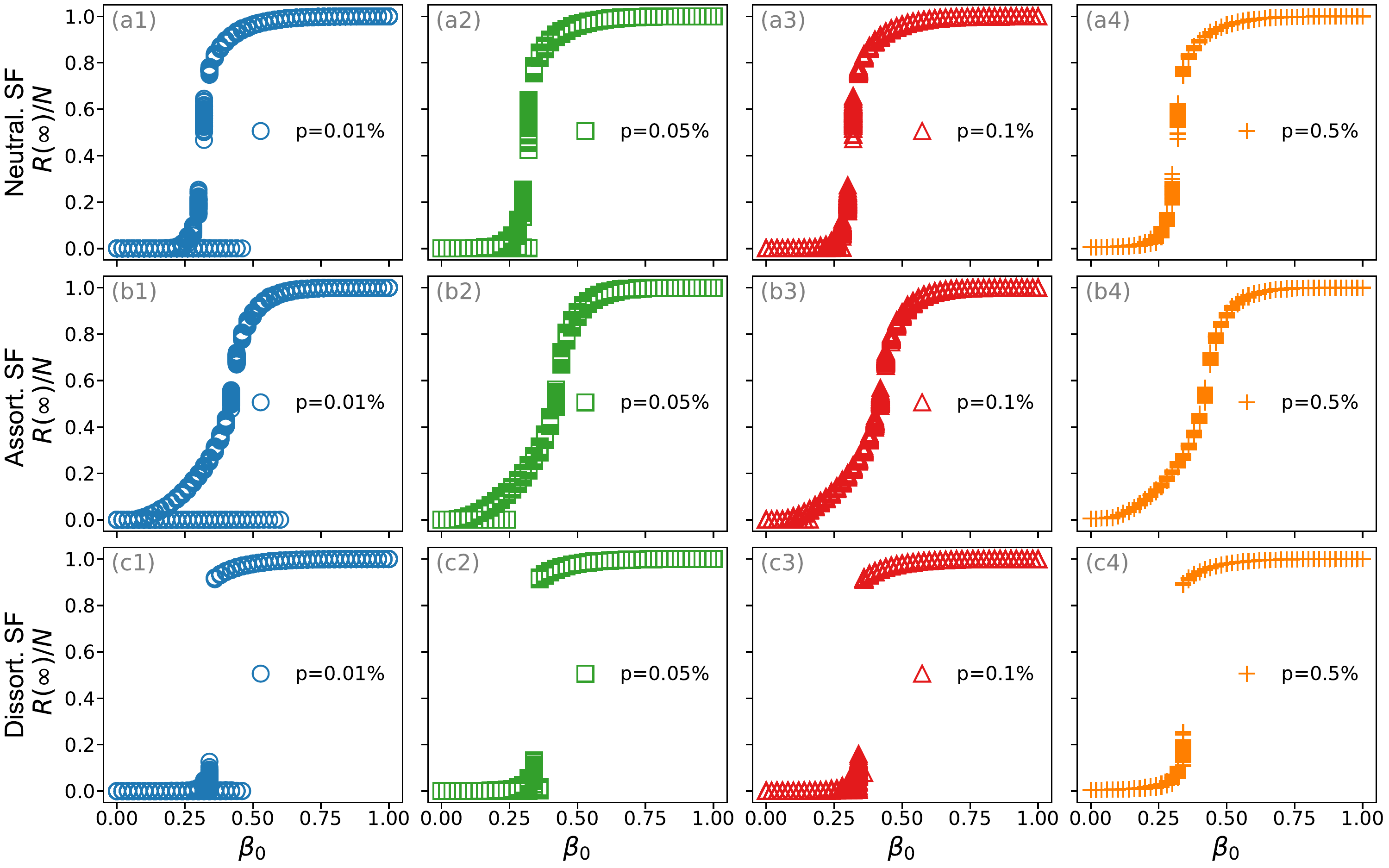}
\caption{
The effect of seed number on the diffusion dynamics.
We show the results on (a1)-(a4) neutral, (b1)-(b4) assortative, and (c1)-(c4) disassortative SF networks, which are the same as those used in Fig.~\ref{fig:Fig3} in the main text. The value of $p$ indicates the percentage of seed nodes in each case. The results are obtained from $100$ independent realizations.
}
\label{fig:Sfig4}
\end{figure*}

\indent
To test the effect of multiple initial seeds on the diffusion process, we consider scale-free networks with different assortativeness levels and vary the number of seeds.
We find that adding more seeds can greatly increase the probability of successfully initiating diffusion (i.e., $R(\infty)/N>5\%$, see Fig.~\ref{fig:Sfig3}) and result into shrinkage of zero-value markers (see Fig.~\ref{fig:Sfig4}), while leaving the dynamical characteristics of the diffusion process unchanged (see Fig.~\ref{fig:Sfig4}). Therefore, we can say the number of seeds does not affect our conclusion from the main text. 

% Create the reference section using BibTeX:
%\bibliography{reference}

\end{document}